\documentclass[fleqn,12pt,twoside]{article}
\usepackage{espcrc1}

\usepackage{graphicx}
\usepackage[figuresright]{rotating}

\newcommand{\AmS}{{\protect\the\textfont2
  A\kern-.1667em\lower.5ex\hbox{M}\kern-.125emS}}

\title{\bf Can Chiral Symmetry Explain the Small Pentaquark Width?}

\author{Dmitri Melikhov and Berthold Stech\\
\vspace{.2cm}
{Institut f\"ur Theoretische Physik der Universit\"at Heidelberg, \\
Philosophenweg 16, 69120, Heidelberg, Germany}%
}
       
\begin{document}

% typeset front matter
\maketitle

\begin{abstract}
It is shown that the decay amplitude for the Jaffe-Wilczek type pentaquarks
is not suppressed by chiral symmetry. On the other hand,  pentaquarks of
positive or negative parity built up of an antiquark and two chirally 
different diquarks in an $S$-state are stable in the limit of a
strictly unbroken chiral symmetry. These states can decay only via the
spontaneous breaking of chiral symmetry. However, this breaking is strong
because of the sizeable magnitude of the quark condensate. Thus, chiral
symmetry cannot be the cause of a tiny decay amplitude, 
even for pentaquarks which are stable in the strict chiral symmetry limit.
\end{abstract}

\vspace{1cm}
%\section{Introduction}
The existence of pentaquarks is not yet undoubtedly established. 
But if these particles exist, the exotic members of the pentaquark
multiplet must have a very small decay width of order 1 MeV or even 
lower \cite{sibirtsev}. 
For the possible origin of the small width 
of the pentaquark many qualitative suggestions have been put forward. 
In a scenario proposed by Jaffe and Wilczek \cite{jw} the pentaquark 
consists of an antiquark and two scalar diquarks a in a relative 
$P$-wave angular momentum state. 
However, a fully dynamical quark model calculation using a non
relativistic Fock-space representation for the pentaquark $\Theta^+$ 
in the Jaffe-Wilczek scenario showed, that the color and 
flavor factors do not reduce the width sufficiently. To get a tiny width   
a small spatial overlap is necessary which requires an uncommon peanut-like
spatial structure of this pentaquark \cite{mss}. 

Recently, the chiral symmetry of QCD was considered as another
possible cause of the small width \cite{beane,ioffe}. 
We examined this point in \cite{ms} by studying the consequences of the broken chiral symmetry 
for the decay amplitude. 

The spontaneously broken chiral symmetry does not leave the vacuum
invariant. Thus, chiral symmetry is not a symmetry of the particle spectrum. Still, 
symmetry arguments can be used for selecting the couplings of fields in effective Lagrangians. 
We therefore consider interpolating 
fields for the usual baryon octet and for the pentaquark and study 
their possible couplings to Goldstone fields. 

We start by defining Lorenz scalar left and right diquark
field operators and state their $SU(3)_L\times SU(3)_R$ 
representations of the chiral group.
Using the convention that the quark fields $q_L$ and $q_R$ transform
according to (3,1) and (1,3), respectively, one finds 
\begin{eqnarray}
D_L^{\alpha,i} =
\epsilon^{\alpha\beta\nu} 
\epsilon^{ijk}
\left((q_L)^T_{\beta,j} 
\gamma_5 C ~(q_L)_{\nu,k}\right), \qquad
D_R^{\alpha,i} =
\epsilon^{\alpha\beta\nu} 
\epsilon^{ijk}
\left((q_R)^T_{\beta,j} 
\gamma_5 C ~(q_R)_{\nu,k}\right).
\end{eqnarray}
Here, $\alpha,\beta,\nu$ are color indices and $C=-i\gamma_0\gamma_2$ is the charge conjugation matrix.
We note that left- and right-handed scalar diquark fields are in
irreducible representation of the chiral group:  
$D_L$ transforms according to $(\bar 3,1)$ and 
$D_R$ according to $(1,\bar 3)$.

Without loss of generality the state vectors of the baryon octet
which contains proton and neutron formed by 3 quarks can be written in a Fock-space representation 
as a quark-diquark combination. It is then suggestive to use among the possibilities
for local baryon field operators an equivalent quark-diquark
form which can generate these baryons 
\begin{eqnarray}
\label{bb}
B_i^j  =  \mbox{$\frac{1}{2}$}\{D_L^{\alpha,j} +D_R^{\alpha,j}  - \gamma_5
(D_L^{\alpha,j} -D_R^{\alpha,j} )\} \;q_{\alpha,i}.
\end{eqnarray}
The baryon field is written in such a way that the left- and right-handed
components are in left and right representations of the chiral group: 
\begin{eqnarray}
\label{form1}
(B_L)_i^j  =   D_L^{\alpha,j}~ (q_L)_{\alpha,i}~, \qquad  (B_R)_i^j  = D_R^{\alpha,j}~ (q_R)_{\alpha,i}~.
\end{eqnarray}
$B_L$ transforms as (1+8,~1), and $B_R$ as (1~,1+8). 

For the interpolating pentaquark field  with two scalar diquarks in a relative $P$-wave state we
take again an expression which leads to separate left and right
representations for $P_L$ and $P_R$:
\begin{eqnarray}
\label {P}
P^{abc} &= &  \mbox{$\frac{1}{2}$}\epsilon_{\alpha\beta\nu}
[( D_L^{\alpha, a}
\stackrel{\leftrightarrow}{ \partial^\mu}
D_L^{\beta,b}+ D_R^{\alpha, a} \stackrel{\leftrightarrow}{ \partial^\mu}D_R^{\beta, b}) 
\nonumber
\\
&-&
\gamma_5(D_L^{\alpha, a}\stackrel{\leftrightarrow}{\partial^\mu}
D_L^{\beta, b}- D_R^{\alpha, a} \stackrel{\leftrightarrow}{ \partial^\mu}
D_R^{\beta,b})]\gamma_5\gamma_\mu C (\bar q^T)^{\nu,c} 
\nonumber\\
P_L^{abc} &=& \epsilon_{\alpha\beta\nu}D_L^{\alpha, a}  
\stackrel{\leftrightarrow}{ \partial^\mu}  D_L^{\beta, b} 
\sigma_\mu i\sigma_2 (q^{*}_L)^{\nu, c}
%(\gamma_5 \gamma_\mu C \bar q_{\gamma,c}^T )_L,
 \nonumber\\
P_R^{abc} &=& \epsilon_{\alpha\beta\nu}D_R^{\alpha, a}  
\stackrel{\leftrightarrow}{ \partial^\mu}  D_R^{\beta, b} 
\bar \sigma_\mu i\sigma_2 (q^*_R)^{\nu, c}.
\end{eqnarray}
Symmetrization with respect to the indices $a$ and $b$ is implied. In (\ref{P}) $P_L, P_R$ and $q_L,q_R$ denote  
two-component Weyl fields transforming as left- and right-handed spinor fields, respectively. 
Obviously, $P$ contains the antidecuplet (with respect to the diagonal subgroup $SU(3)_V$) we are interested in and can be extracted from it. The parity of $P$ is even. The expression for $P$
includes (covariant) derivatives to represent the relative 
$P$-wave state of the two diquarks in a local form.
Evidently,  
$P_L$ transforms as $(8+\overline{10},1)$ and $P_R$ as $(1,8+\overline{10})$. Consistent with the chosen form
for the baryon octet all quarks in $P_L$ are left-handed and all
quarks forming $P_R$ are right-handed.

The combination 
$B^{\dagger}_L \overline\sigma_\mu  P_L$, 
can form a left-handed 
octet current transforming as $(8,1)$ when applying proper Clebsch-Gordan coefficients. 
This is evident from the transformation properties given above.
Similarly,
$B^{\dagger}_R  \sigma_\mu P_R$
can form a right-handed current transforming as $(1,8)$. 
Together, these combinations have the correct properties of an axial vector 
octet and therefore can couple to an axial vector field with a 
chirally invariant coupling constant! 

Therefore, one finds a finite axial vector coupling constant and can
construct a chirally invariant derivative coupling of pseudoscalar mesons
for the pentaquark to nucleon transitions.
There exists no symmetry
argument for the corresponding coupling constant to vanish. This
result is in agreement with the numerical values for the width
obtained in \cite{mss} which turned out to be generally large. Only a
small spatial overlap between this pentaquark and the nucleon can
reduce the width. The existence of a chirally invariant coupling for
this pentaquark can be traced back to the fact that $\gamma_\mu C \bar
q^T$ transforms like the diquarks. Without the $\gamma_\mu $ term and thus without 
the $P$-wave structure the result will be quite different.

%\vspace{1cm}
Therefore, let us now consider an interpolating pentaquark field
of positive parity which generates the two diquarks in an $S$-wave state:
\begin{eqnarray}
\label{ps}
P^{abc} &=&  \mbox{$\frac{1}{2}$}\epsilon_{\alpha\beta\nu}[(D_L^{\alpha, a}
D_R^{\beta,b}+D_R^{\alpha, a}  D_L^{\beta, b}
%\nonumber\\
%&&
- \gamma_5 (D_L^{\alpha, a}
D_R^{\beta,b}-D_R^{\alpha, a}  D_L^{\beta,b})]~\gamma_5 C~  (\bar q^T)^{\nu,c}.
%\nonumber\\
\end{eqnarray}
In the two-component Weyl field representation  we have
\begin{eqnarray}
\label{PLR}
P_L^{abc}  &=&  \epsilon_{\alpha\beta\nu}~D_L^{\alpha, a}  D_R^{\beta, b}  
%C ~((\bar q_{\gamma,c})^T)_L~,
i \sigma_2 (q_R^*)^{\nu,c},
\nonumber
\\
P_R^{abc}  &=&   \epsilon_{\alpha\beta\nu}~D_R^{\alpha, a}  D_L^{\beta, b} 
i \sigma_2 (q_L^*)^{\nu,c}.  
%C ~((\bar q_{\gamma,c})^T)_R~.
\end{eqnarray} 
$P_L$ transforms as $(\bar 3 , 3 + \bar 6)$ and $P_R$ as $(3 + \bar 6 ,  \bar 3)$. 
The pentaquark $SU(3)_V$ antidecuplet arises from the $\bar 6$
content in these expressions.

The interpolating field operator which generates $S$-wave pentaquarks of negative parity is obtained by multiplying (\ref{ps})
by $\gamma_5$. Evidently, this negative parity pentaquark has the same transformation properties under chiral transformation as the one with positive parity.

As it is seen from these transformation properties, this time 
the left-handed axial current formed from $B_L^\dagger P_L$ transforms as 
$(\bar 3+6+\bar 15,3+\bar 6)$ and not as $(8,1)$ as required. This is in strong contrast to the case of the
"$P$-wave" type pentaquark we discussed before. 

Because of the transformation properties of the $S$-wave pentaquarks it is also not possible to construct 
their invariant coupling to the Goldstone particles. 
Therefore, the pentaquark with the $S$-wave structure is stable in the fully unbroken chiral limit 
but can decay with an amplitude proportional to the quark condensate from the spontaneous symmetry breaking. 
The stability of the "$S$-wave pentaquarks"
is in accord with the findings of Ioffe and Oganesian \cite{ioffe}. 

According to the derivation it is clear that the precise internal structure is not essential, 
only the transformation properties matter: The diquarks do not have to be of a small 
size and may strongly overlap with each other and the antiquark.
 
The $P$-wave pentaquark and the positive parity $S$-wave pentaquark have identical quantum numbers: 
total angular momentum, $SU(3)$ quantum numbers, and parity. 
But they differ in their chiral transformation properties and their $\gamma_5$-parity 
(a discrete subgroup of $U(3)_L \times U(3)_R$): multiplying all quark fields by $\gamma_5$ gives 
$+\gamma_5$ for the "$P$-wave pentaquark" and the nucleon octet but $-\gamma_5$ 
for the $S$-wave pentaquark. The axial vector constructed from the baryon octet and the 
latter pentaquark then changes sign under this transformation \cite{ioffe}, another reason 
for a vanishing coupling to the axial field in the strict chiral limit.

%****************************************************************************

The spontaneous breaking of chiral symmetry however changes the situation \cite{ms}. 
Quark condensates appear and the light meson octet,
the Goldstones, can be represented in the non-linear form 
\begin{eqnarray}
\label{sigma}
\Sigma  =  \exp \left(i \lambda_k \phi^k/{F_\pi}\right). 
\end{eqnarray}
The unitary matrix $\Sigma$ transforms according to $(3,\bar 3)$. 
Here $\lambda^k$ denote the Gell-Mann matrices and $\phi_k$ the 8 pseudoscalar 
meson fields. Obviously, chiral transformations of $\Sigma$ leave
this matrix unitary and define thereby the transformation properties of the
Goldstone fields. The matrix $\Sigma$ can now be used to change the transformation
properties of the fields: 
$\Sigma q_R$ transforms as $q_L$, and $\Sigma^\dagger q_L$ 
transforms as $q_R$. An appropriate application of $\Sigma$
allows the non vanishing of the axial vector matrix element for the transition 
to nucleons.
Thus, because of the spontaneous symmetry breaking also the $S$-wave  
pentaquark looses its stability.
In a Fock-space model where all quark fields act on the vacuum at equal 
times (or on a lightlike hyper-plane) but at different positions it is 
easy to see the reason for the stability of the $S$-wave pentaquark in 
case of the strict chiral symmetry, and for its instability due to the 
spontaneous chiral symmetry breaking: In the unbroken case the axial vector 
current matrix element for the transition amplitude can be calculated by 
commuting the fields using the equal time commutation relations. As it is 
obvious from (\ref{form1}) and (\ref{PLR}) this gives zero for our 
$S$-wave pentaquark since $q_L$ commutes with $q^{\dagger}_R$ and $q_R$ with
$q^{\dagger}_L$. However, since chiral symmetry is spontaneously broken, 
nonlocal condensates such as $\langle \psi_L(x) \psi^{\dagger}_R(0)\rangle|_{x_0=0} $ 
survive. (The correct gauge invariant form for these condensates includes 
a Schwinger string not  shown here.) These condensates replace the 
$\delta^3(x)$ function obtained from equal time commutators in transitions 
which are not suppressed by chiral symmetry. One can compare now  the space 
integral of the condensate with the space integral of the $\delta^3(x)$ 
function (which is 1). This gives a measure of the importance of the 
spontaneous symmetry breaking. Taking \cite{cond}
\begin{equation}
\langle\psi_L(x_0=0,\vec x) \psi^{\dagger}_R (0)\rangle= 
\frac{1}{2} \langle \bar \psi \psi \rangle e^{-\vec x^2 M^2_0/16}
\end{equation}
with $\langle\bar \psi \psi\rangle \approx (254~{\rm MeV})^3$ and $M_0
\approx 860~{\rm MeV}$, 	 
the numerical value of the space integral turns out to be $\approx 4.6$.
In an actual calculation of the transition amplitude this space
integral will be somewhat reduced by the variation of the wave
function multiplying the condensate, but it will certainly stay of
order one. Consequently, we are forced to conclude: In spite of
the vanishing of the decay amplitude in the unbroken chiral symmetry
limit, the spontaneous breaking of this symmetry is strong and leads 
in general to amplitudes comparable with the ones which are not
inhibited by the unbroken symmetry.

We have seen that 
{\bf chiral symmetry cannot be responsible for the small pentaquark width}. 
If pentaquarks exist their small width must have a
different origin. The most likely cause is an unusual spatial structure of
these particles leading to a small wave function overlap with the nucleon
wave function \cite{mss}.

\end{document}